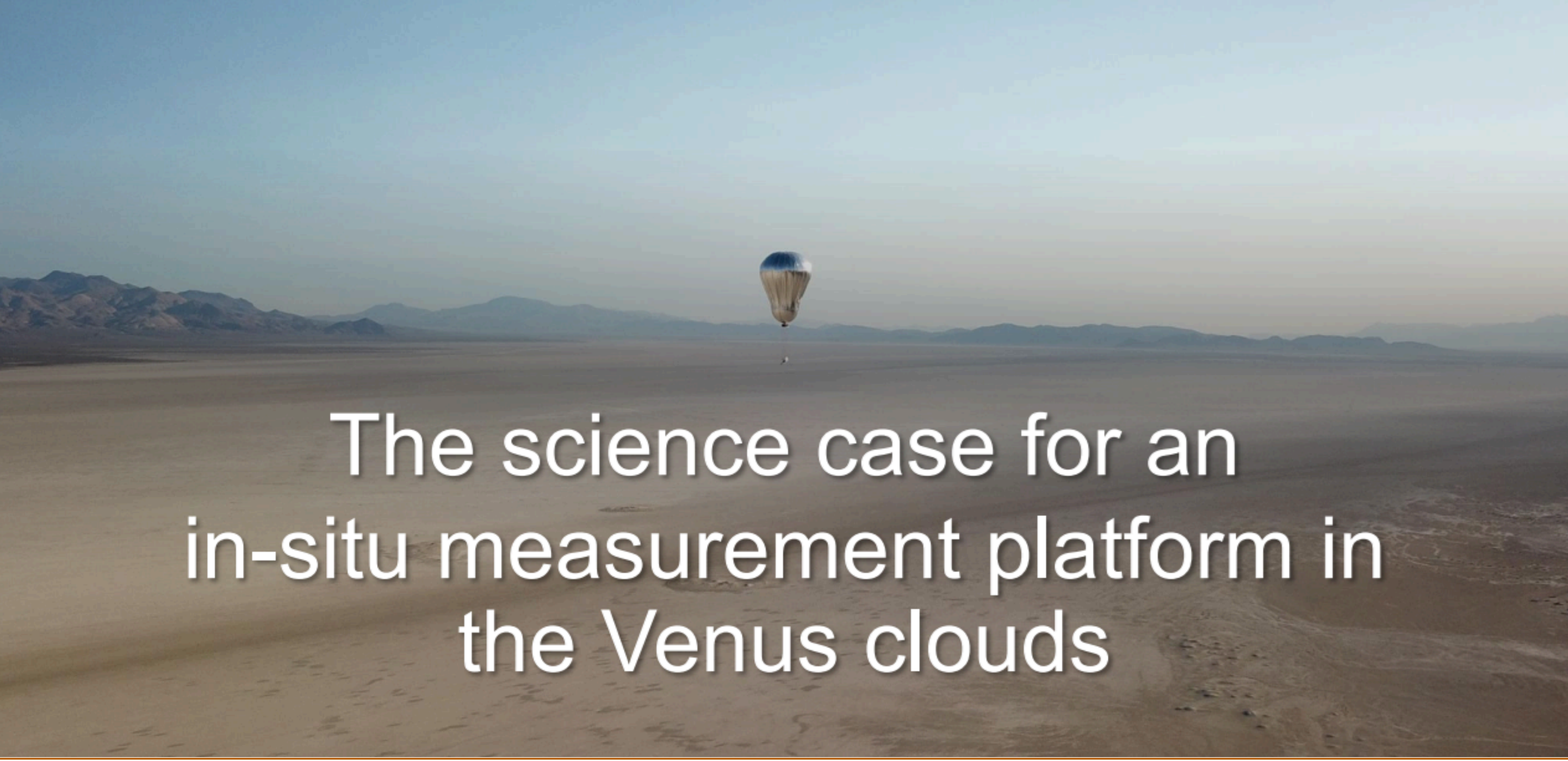

# The science case for an in-situ measurement platform in the Venus clouds

*Image credit: Izraelevitz et al. 2024*

**Lead authors:**

Maureen Cohen (The Open University)
maureen.cohen@open.ac.uk

Jo Egan (The Open University)

**Co-authors:**

Martin Airey (University of Reading)
Karen Aplin (University of Bristol)
Gilles Bailet (University of Glasgow)
Paul Byrne (Washington University in St Louis)
Chris Castelli (The Open University)
Charles Cockell (University of Edinburgh)
Nathan Mayne (University of Exeter)
Blair McGinness (University of Reading)
Ingo Mueller-Wodarg (Imperial College London)
John Plane (University of Leeds)
Julia Semprich (The Open University)
Denis Sergeev (University of Bristol)
Wencheng Shao (University of Southampton)
Alexander Stott (Institut Supérieur de l'Aéronautique et de l'Espace)
Roland Young (University of Aberdeen)

**Co-signers:**

Jo Barstow (The Open University)
Ananyo Bhattacharya (University of Michigan)
Valentin Bickel (University of Bern)
Marrick Braam (University of Bern)
Megan Brown (University of Cambridge)
Alejandro Cardesín (Aurora/SSC Space)
Dave Clements (Imperial College London)

James Cutts (Jet Propulsion Laboratory, California Institute of Technology, ret.)
Benjamin Frandsen (University of Helsinki)
Joshua Ford (University of Bristol)
Richard Ghail (Royal Holloway University of London)
Gabriella Gilli (Instituto de Astrofísica de Andalucía)
Jane Greaves (University of Cardiff)
Nora Haenni (University of Bern)
James Holmes (The Open University)
Anna Horleston (University of Bristol)
Kyle Horn (Massachusetts Institute of Technology)
Jessica Irving (University of Bristol)
Patrick Irwin (University of Oxford)
Sean Jordan (ETH Zurich)
Shubham Kulkarni (University of Oxford)
Mark Lee (The Open University)
Franck Lefèvre (Sorbonne Université)
Maxence Lefèvre (Sorbonne Université)
Dan Li (Université de Versailles Saint-Quentin)
Arnaud Mahieux (Aurora/SSC Space)
Julia Maia (German Aerospace Center)
Emmanuel Marcq (LATMOS / Université de Versailles Saint-Quentin)
Philippa Mason (Imperial College London)
João Mendonça (University of Southampton)
Jan-Peter Muller (University College London Mullard Space Science Laboratory)
Brian P. Murphy (University of Edinburgh)
Claire Newman (Aeolis Research)
Kevin Olsen (University of Oxford)
Manish Patel (The Open University)
Michael Pauken (Jet Propulsion Laboratory, California Institute of Technology)
Moa Persson (Swedish Institute of Space Physics)
Janusz Pętkowski (Politechnika Wrocławska)
Arianna Piccialli (Royal Belgian Institute for Space Aeronomy)
Ben Rider-Stokes (The Open University)
Andreas Riedo (University of Bern)
Paul Rimmer (University of Cambridge)
Sara Seager (Massachusetts Institute of Technology)
Aurélien Stolzenbach (Instituto de Astrofísica de Andalucía)
Nicolas Streel (Institute for Basic Science)
Paul Streeter (The Open University)
Paola Tiranti (Northumbria University)
Ann Carine Vandaele (Royal Belgian Institute for Space Aeronomy)
Iris van Zelst (University of Edinburgh)
Audrey Vorburger (University of Bern)
Peter Wurz (University of Bern)

*Acknowledgements: The authors would like to thank Colin Wilson for his expert advice.*

**Executive summary:** The next step for Venus exploration after the remote sensing missions of the early 2030s will be in-situ aerial platforms, the Venus equivalent of rovers for Mars. Sophisticated Venus balloon technology is already under development and presents an opportunity for the United Kingdom to act as a bilateral partner in missions exploring key questions about the volcanic and geological history of Venus, the potential habitability of the Venus clouds, and Venus as an analogue for exoplanets.

**I. Introduction:** Venus exploration has seen a surge of renewed interest. Often called Earth's "sister planet," Venus is similar in its size, mass, and distance from the Sun, yet its atmosphere and climate are vastly different. Researchers have sought to determine if Venus ever supported liquid water, how its climate has developed over time, and the nature of its volcanism. As exoplanet science has expanded and increasing numbers of terrestrial planets are found within the inner edge of the habitable zone, Venus has drawn attention as an exoplanet analogue and *de facto* laboratory for studying planets with post-runaway greenhouse climates, slow rotation periods, non-water-based clouds, or superrotating atmospheres. Finally, the disputed report of a detection of a biosignature gas, phosphine, on Venus in 2020 has reignited a historic discussion about the habitability of the Venus clouds–the only place in the solar system with temperatures and pressures comparable to Earth's surface–as well as the role of biosignature claims in the search for extraterrestrial life inside and outside the solar system.

The NASA missions VERITAS[1] and DAVINCI[2] and the ESA mission EnVision are now slated to visit Venus in the early 2030s. These missions will return a wealth of remote sensing data about the planet's surface, its atmosphere above the cloud tops, and select regions of the deep atmosphere observable through spectral windows. The DAVINCI mission includes a probe intended to sample the atmosphere during a brief descent to the surface. Yet these remote sensing-centred missions have intrinsic limitations on cloud-covered Venus; in fact, despite decades of observation, including long-term remote sensing missions like the Japanese Akatsuki Venus Climate Orbiter and ESA's Venus Express, our understanding of the Venus cloud environment, deep atmosphere, and surface–atmosphere interactions remains extremely limited. Major questions remain:

- Why does the sulphur dioxide content drop precipitously across the cloud decks? Is sulphur dioxide variability caused by volcanic activity?
- What are the rapidly shifting UV-absorbing species seen in the upper clouds?
- Does convection in the Venus clouds imply storms and the presence of lightning?
- Does the cloud environment meet the definition of habitability? Are the clouds in chemical disequilibrium and, if so, why?
- How do the surface and the atmosphere interact? Are chemical species in the atmosphere sourced from the surface?
- How is Venus' geological history recorded in its atmospheric composition and structure?

There is recognition in the global Venus community that advancing not only Venus science, but planetary science, exoplanet climatology, and habitability theory requires long-term, in-situ measurements on Venus. Development of Venus cloud-level balloon concepts to meet this need is already underway. In this white paper, we outline the science case for the United Kingdom's involvement in and potential contributions to such a mission.

[1] Venus Emissivity, Radio science, InSAR, Topography And Spectroscopy
[2] Deep Atmosphere Venus Investigation of Noble gases, Chemistry and Imaging

**II. Key science questions:**

*Volcanism*: How volcanically active is Venus? The nature of Venus' geological activity remains an open question. Little is known about the internal structure, interior–surface coupling, volcanism, and possibility of quakes on Venus. Atmospheric studies can reveal details of both present-day active volcanism and tectonics (Garcia et al. 2024, Wilson et al. 2024, Airey et al. 2015) and the interior structure and evolution of the planet (Weller et al. 2023) through measurements of volcanic trace gases and atmospheric waves triggered by seismic activity. A better understanding of Venus provides a key contrast with the Earth, which can help elucidate how rocky planets evolve over geological time.

*Habitability*: Are the Venus clouds habitable? Venus is uniquely Earth-like in the solar system as the only other body with room-temperature and -pressure conditions. These conditions occur in the Venusian clouds at 47–70 km altitude, making them a key location for comparative study to the Earth (Cockell 1999). Reported detection of "biosignature" molecules (Greaves et al. 2020, 2021a, 2021b) demonstrates the need for improved understanding of the conditions and habitability of the Venusian clouds (Patel et al. 2022, Dartnell et al. 2015). If Venus was ever habitable in the past, its atmosphere may also preserve evidence of this in isotope ratios. A comprehensive understanding of how the atmospheric and aerosol composition varies in the cloud region and whether lightning discharges occur (McGinness 2024, Sukanjanajtee & Aplin 2024) are vital to improve understanding of how the region is simultaneously similar and different to Earth's atmosphere. Even if the clouds are uninhabitable, they can provide important scientific information on what factors drive a terrestrial rocky planet and its atmosphere away from habitable conditions and which chemical species are barriers to life.

*Exoplanet science*: Can Venus help us learn about exoplanets? Beyond the Earth, Venus' similarity to detected rocky exoplanets makes it an ideal analogue for the study of the atmospheres of slowly-rotating or tidally locked planets (Cohen et al. 2024, 2025), providing a unique opportunity for the study of atmospheric transport and dynamics in exoplanet atmospheres. As many known terrestrial exoplanets orbit within the inner edge of the circumstellar habitable zone, Venus may also be representative of these hot climates (Parkinson et al. 2025), and understanding its photochemistry can help us interpret observational data from potentially similar atmospheres (Jordan et al. 2021)

**III. Research opportunities:**

a. *Atmospheric composition and chemistry*

The Venusian cloud region is photochemically active, with the chemical cycles of many species currently not understood. Well-developed numerical models are available for atmospheric retrievals and the historic and present-day climate (e.g. Irwin et al. 2008, Perez-Hoyos et al. 2018, Giles et al. 2022, Cohen et al. 2024, 2025, Constantinou et al. 2025, Egan et al. 2025a), but the scarcity of recent observations in and below the cloud deck limits our ability to constrain atmospheric coupling and transport processes. Equipping an in-situ measurement platform with a mass spectrometer to measure gas-phase concentrations of various species would provide measurements of the variability of key chemical compounds in the cloud layer. For example, the cause of the depletion of $SO_2$ through the cloud deck (Vandaele et al. 2017) could be a combination of variability in below-cloud $SO_2$, perhaps from intermittent volcanism (Wilson et al. 2024), formation of other sulphur-containing reservoir species (e.g. Pinto et al. 2021, Frances-Monerris et al. 2022), or increased uptake into cloud droplets (Rimmer et al. 2021, Jiang et al. 2024, Ubukata et al. 2025), and in-situ measurements over a period of time would aid in understanding the contribution of each pathway. Measurements or upper limits of other species, such as phosphine, or other phosphorus-containing species, would provide clarity on the recent controversial biomarker observations (Greaves et al. 2020, 2021a, 2021b), or could provide indications of novel (abiogenic) chemistry in the Venus sky.

b. *Aerosol science:*
The formation, growth, and loss of Venusian cloud particles is not well understood, with the composition and size distributions of the multimodal cloud deck broadly unmeasured since the 1978 Pioneer Venus probes, and uncertainty in cloud composition limiting microphysical modelling of the clouds (McGouldrick, 2022, Karyu et al. 2025). Aerosol particle-sizing technology has improved markedly in the intervening decades, and measurements of Venusian cloud particle sizes would greatly increase understanding of sulphur and other chemical cycles on Venus and, by extension, the Junge layer on Earth. Atmospheric electricity can influence the formation of cloud droplets and their interaction with chemical species (McGinness 2025, Airey et al. 2024, Aplin 2006). Instruments such as a relaxation probe or loop antenna could measure atmospheric electricity and its impacts on the atmosphere for the first time.

As the habitability of the cloud deck is a key consideration, measuring the acidity, salinity, and CHNOPS content of the sulphuric acid cloud droplets is vitally important. The presence of additional dissolved or suspended particulate species in the cloud droplets, such as salts to buffer the pH (Rimmer et al. 2021) or products of the reactions of meteoric metals, may provide reservoirs for additional $SO_2$ uptake into the cloud droplets, and are leading candidates for the cause of the highly variable UV absorption (Jiang et al. 2024, Egan et al. 2025b). Quantification of the composition of the cloud droplets by an aerosol mass spectrometer would provide the first in-situ cloud aerosol composition measurements ever, and identification of UV-absorbing or sulphur-containing species in the droplets could provide answers to several key questions about the chemical cycles and the radiation balance within the clouds.

c. *Atmospheric dynamics:*
A long-term measurement platform could detect 4–5-day-period planetary-scale waves, gravity waves, and solar thermal tides thought to contribute to the superrotation of the Venus atmosphere, a long-standing topic of inquiry (Imamura et al. 2020, Sánchez-Lavega et al. 2017). Previously, such waves could be detected at the cloud tops but not within the clouds themselves, leaving large gaps in understanding due to the intrinsically three-dimensional nature of the superrotation mechanisms. Wave-driven superrotation is also predicted by atmospheric modelling of exoplanets, with potentially similar drivers, so detailed characterisation of the Venus general circulation and wave population additionally validates such model predictions and supports exoplanet science (Cohen et al. 2025, Cohen et al. 2024).

The lowest Venus cloud deck from ~47–55 km is convectively active. Convective plumes may drive the observed short-term variability in sulphur dioxide content at the cloud tops and control mixing of chemical species transported from the deep atmosphere with those produced photochemically or delivered through meteoric dust above the clouds (Lefèvre et al. 2024, Lefèvre et al. 2022, Lefèvre et al. 2018). In addition, convective activity could cause storms, including lightning (Constantinou et al. 2025). A direct detection of lightning on Venus would be a first and have implications for the chemical composition and equilibrium state of the atmosphere. Atmospheric electricity instruments or a magnetometer could detect lightning, while anemometers could register the frequency and strength of convective plumes, as was briefly demonstrated in the Vega balloon missions (Linkin et al. 1986).

Venus climate modelling is advanced, with well-developed open-source models available (Shao et al. 2024, Mendonça & Buchhave 2020, Lebonnois et al. 2010). High-quality, long-term measurements can be incorporated into Venus atmospheric modelling through data assimilation as has previously been done for Mars (Young et al. 2022, Holmes et al. 2020). Reanalysis datasets that include the upper atmosphere, cloud decks, deep atmosphere, and coupling between these different regions would facilitate the establishment of a true Venus climatology, enabling investigation of, for instance, long-term variability (Marcq et al. 2013, Esposito 1984).

d. *Surface–atmosphere interaction:*
Finally, pressure sensors on a cloud-level balloon platform could detect infrasound waves generated by seismic activity at the Venus surface (Garcia et al. 2024, Garcia et al. 2022). This would be the first detection of venusquakes and a crucial missing element of our understanding of Venus' interior and geodynamics. Balloon-based chemistry and aerosol sensors as described above could seek impacts of volcanic eruptions in the clouds.

**Table 1**: *Summary of research objectives, measurements, and possible instruments*

| **Objective** | **Measurement** | **Instrument** |
|---|---|---|
| Trace the impact of volcanism on the Venus atmospheric composition | Volcanic trace gas abundances | Mass spectrometer |
| Reconstruct planetary and climate evolution by constraining water budget | Noble gases and isotope ratios | " |
| Shed light on novel chemistry, chemical disequilibria, and habitability | Other key chemical species, e.g. CHNOPS | " |
| Understand physics, chemistry, and formation of non-water-based clouds, including heterogeneous chemistry | Composition and size of cloud and haze particles | Aerosol particle sizing instrument/aerosol mass spectrometer |
| Detect lightning and atmospheric | Atmospheric electricity | Relaxation probe or |
| electricity | | loop antenna |
| Detect lightning | Magnetic fields | Magnetometer |
| Measure convective plume activity, detect planetary-scale waves, solar tides, and gravity waves | Wind velocities | Anemometer and IMU (inertial measurement unit) |
| Detect quakes and meteor impacts | Pressure waves | Pressure sensor |

**IV. Technological & strategic context:**

A Venus cloud-level balloon offers advantages compared to other mission concepts. Unlike a descent probe, a balloon enables long-duration observations, staying aloft for weeks or months and providing continuous data on weather and climate. A balloon would explore rarely studied parts of the atmosphere, but is significantly easier and cheaper to engineer than a surface lander. Aerial platforms on Venus are analogous to rovers on Mars and the next logical step in Venus exploration beyond remote sensing.

The Soviet Union successfully deployed two balloons carrying pressure and temperature sensors, anemometers, and nephelometers into the Venus cloud decks at 54 km altitude as part of the 1986 Vega missions. Over 46 hours, these platforms provided invaluable data about the cloud environment. Super-pressure balloons like the Vega balloons are at a relatively high technological readiness level due to successful past use on Venus. Variable altitude control, corrosion protection for a long-term stay, and other aspects such as deployment require further work; however, no “technological show-stoppers” impede the development of these capabilities (VexAG Venus Technology Plan 2019, Cutts et al. 2018). Enormous improvements in scientific ballooning technology, including in autonomy, material science, and miniaturisation, have made a longer-lasting and more sophisticated balloon platform feasible and garnered substantial support from the international Venus community.

Potential routes for UK partnership in a Venus balloon mission include:

1. *JPL*: The Jet Propulsion Laboratory at California Institute of Technology is developing a variable-altitude long-term Venus balloon and has successfully completed prototype test flights in the United States (Byrne et al. 2025, Izraelevitz et al. 2024, Byrne et al. 2023). Research at the Massachusetts Institute of Technology is exploring in-situ resource utilisation to generate propellant and extend the mission lifetime for months or even years (Horn et al. 2025). JPL operates a Planetary Cloud and Aerosol Research Facility (Pauken et al. 2025, Pauken et al. 2023) which allows for preparatory experimental work for a Venus cloud platform. The UK could provide an instrument or contribute a technical sub-system to a JPL-led mission if selected by NASA.
2. *MIT:* The Venus Morning Star missions are a privately funded series of missions led from MIT that include an aerial platform concept to explore the potential habitability of the clouds (Seager et al. 2021). The first launch, the Rocket Lab Mission to Venus, constitutes a descent probe only and is planned for 2026. The second mission, prospectively in 2030 or 2031, will include a balloon or parachute. The University of Exeter and Open University have already provided scientific advice and modelling support for the 2030s mission. The UK could expand its involvement and also provide an instrument.

3. *ESA:* Extensive scoping of a balloon mission was carried out for the European Venus Explorer concept submitted to ESA in 2010 as an M-class mission (Wilson et al. 2012, Chassefière et al. 2009a, Chassefière et al. 2009b). Teams in other ESA member states are also developing balloon proposals (Wieser et al. 2023, Garcia et al. 2024, 2025). The UK could carry out a roadmapping exercise to identify technologies, partners, and pathways to an M-class ESA Venus balloon mission.

Other possibilities on the 2040s horizon could include a contribution to an Indian Space Research Organisation mission, as ISRO is currently planning a Venus orbiter. Given widespread support in the international Venus community for aerial platforms as the next step in Venus exploration, we believe the United Kingdom should position itself as a partner in these upcoming developments.

The UK Venus community has expertise in Venus atmospheric chemistry (The Open University, University of Leeds, University of Cambridge), atmospheric electricity and lightning (University of Bristol, University of Reading, University of Cambridge), Venus climate modelling (The Open University, University of Exeter, University of Southampton, University of Bristol, University of Oxford), data assimilation for planetary atmospheres (The Open University, University of Oxford, University of Aberdeen), and astrobiology, including Venus-specific (University of Edinburgh, University of Cardiff, Imperial College London). The UK has a strong representation in instrumentation for space missions and could contribute a mass spectrometer, magnetometer, or technical subsystem such as a communications or attitude and position determination system. Thanks to its mature space industry, the UK also offers emerging space technologies such as femtoprobes (University of Glasgow) and self-dispersing anemometers (SpaceAM, University of Exeter) which may reach technological readiness on a timescale suitable to a Venus balloon mission.

**V. Conclusion & recommendations:**

An in-situ measurement platform in the Venus clouds is a promising mission candidate in the coming decade, with significant technological readiness, international support, and scientific importance. The UK planetary science community should position itself for involvement in such a mission through contribution of instrumentation, technical subsystems, data processing, and/or scientific expertise. Concrete enablers of a Venus aerial platform mission include:

1. Sustaining and growing the community through:
    a. Establishment of a UK-wide Venus science meeting or coordination group
    b. Support for UK involvement in current Venus science, in particular EnVision
2. Developing future bilateral opportunities on the 2035/2040 horizon through:
    a. Relationship-building with potential mission partners such as JPL and MIT
    b. Commissioning of a mission technical feasibility study to establish roadmap towards an ESA M-class mission or participation in a Venus Morning Star mission
    c. Identification of and investment in key opportunities for UK involvement, e.g. a mass spectrometer, miniaturised sensor system, data assimilation codebase